%% file: main.tex
\documentclass[final,twoside,11pt]{entics}
\pdfoutput=1
\usepackage{enticsmacro}
\usepackage[utf8]{inputenc}
\usepackage{newunicodechar}
\usepackage{microtype}
\usepackage{bold-extra}
\usepackage[shortlabels]{enumitem}

\usepackage{booktabs}
\usepackage{subcaption}
\usepackage{listings}
\lstdefinelanguage{JuliaMin}%
  {morekeywords={baremodule, begin, break, catch, const, continue, do, else, elseif, end, export, false, finally, for, function, global, if, import, let, local, macro, module, quote, return, struct, true, try, using, while,
                 @theory, @theorymap, @instance, @withmodel},%
   sensitive,
   morecomment=[l]\#,%
   morestring=[b]"%
 }[keywords,comments,strings]

\usepackage{gatlab-macros}
\usepackage{cleveref}

\newunicodechar{⊣}{$\dashv$}
\newunicodechar{Γ}{$\Gamma$}
\newunicodechar{⋅}{$\cdot$}
\newunicodechar{⊗}{$\otimes$}
\newunicodechar{∉}{$\notin$}
\newunicodechar{ℕ}{$\mathbb{N}$}

\def\conf{MFPS 2024}
\def\myconf{mfps40} 	
\volume{4}			
\def\volu{4}			
\def\papno{17}			
\def\mydoi{14666}%
\def\lastname{Lynch, Brown, Fairbanks, Patterson}
\def\runtitle{GATlab}
\def\copynames{O. Lynch, K. Brown, J. Fairbanks, E.}
\def\copyname{\quad Patterson\quad }

\begin{document}

\begin{frontmatter}
  \title{GATlab: Modeling and Programming with Generalized Algebraic Theories}
  \author{Owen Lynch\thanksref{topos}\thanksref{olmail}}
  \author{Kris Brown\thanksref{topos}\thanksref{kbmail}}
  \author{James Fairbanks\thanksref{uf}\thanksref{jfmail}}
  \author{Evan Patterson\thanksref{topos}\thanksref{epmail}}
  \address[topos]{Topos Institute, Berkeley, CA}
  \address[uf]{University of Florida,
    Computer \& Information Science \& Engineering,
    Gainesville, FL}
  \thanks[olmail]{Email: \href{mailto:owen@topos.institute}%
    {\texttt{\normalshape owen@topos.institute}}}
  \thanks[kbmail]{Email: \href{mailto:kris@topos.institute}%
    {\texttt{\normalshape kris@topos.institute}}}
  \thanks[jfmail]{Email: \href{mailto:fairbanksj@ufl.edu}%
    {\texttt{\normalshape fairbanksj@ufl.edu}}}
  \thanks[epmail]{Email: \href{mailto:evan@epatters.org}%
    {\texttt{\normalshape evan@epatters.org}}}

  \begin{abstract}
    Categories and categorical structures are increasingly recognized as useful
    abstractions for modeling in science and engineering. To uniformly implement
    category-theoretic mathematical models in software, we introduce GATlab, a
    domain-specific language for algebraic specification embedded in a technical
    programming language. GATlab is based on generalized algebraic theories
    (GATs), a logical system extending algebraic theories with dependent types
    so as to encompass category theory. Using GATlab, the programmer can specify
    generalized algebraic theories and their models, including both free models,
    based on symbolic expressions, and computational models, defined by
    arbitrary code in the host language. Moreover, the programmer can define
    maps between theories and use them to declaratively migrate models of one
    theory to models of another. In short, GATlab aims to provide a unified
    environment for both computer algebra and software interface design with
    generalized algebraic theories. In this paper, we describe the design,
    implementation, and applications of GATlab.
  \end{abstract}

\end{frontmatter}

\input{01-introduction.tex}
\input{02-background-gats.tex}
\input{03-gatlab-gats.tex}
\input{04-background-gat-morphisms.tex}
\input{05-applications.tex}

\begin{ack}
  The authors acknowledge support from the DARPA ASKEM program under grant
  HR00112220038.
\end{ack}

\bibliographystyle{./entics}
\bibliography{gatlab}

\appendix

\input{A1-implementation.tex}

\end{document}

%% file: 01-introduction.tex
\section{Introduction}

Category theory has long been recognized as a useful organizing principle for
programming. The far-ranging dictionary between category theory and type theory,
originating with the equivalence between cartesian closed categories and lambda
calculus with product types \cite{lambek_Introduction_1986}, has led language
designers to import all manner of categorical concepts into programming
languages. In this usage, category theory acts a \emph{mathematical model of the
  programming language}, often giving its denotational semantics. But this is
not the only role that category theory can play in programming. Applied category
theorists have shown how category theory can formalize compositional structure
latent in science and engineering, from relational databases to stochastic and
quantum processes to differential equations and dynamical systems \cite{spivak_Functorial_2012,fong_Causal_2013,coecke_Categorical_2018,libkind_Operadic_2022}. Category
theory now becomes a \emph{mathematical model of the subject-matter domain},
which (like any model) will be most useful if it can be implemented in software.
We thus need a software system that can readily express and compute with
categorical structures.

This objective might be served by designing the host language according to
category-theoretic ideas but is in principle independent of it. Indeed, there is
presently a tension between the two roles, since the dependently typed languages
that most natively express categorical concepts tend to have few facilities for
scientific and technical computing, whereas the languages most commonly used by
scientists and engineers tend to have minimalistic type systems. The problem can
be approached from either direction. In this work, we show how to augment an
existing technical computing language with an advanced type system to enable
uniform computation over categorical structures.

Rather than trying to retrofit a full dependent type theory onto a loosely
typed programming language, we work with generalized algebraic theories, an
extension of algebraic theories sufficient to axiomatize categorical
structures. A \emph{generalized algebraic theory} (GAT)
\cite{cartmell_Generalised_1978,cartmell_Generalised_1986}, more suggestively
called a \emph{dependently typed algebraic theory}
\cite{pitts_Categorical_2001}, is similar to a (typed or sorted) algebraic
theory except that its types may depend on terms. The ur-example of a GAT is
the theory of categories, in which the type of a morphism depends on a pair of
objects, the domain and codomain. Though a relatively simple type theory, GATs
suffice to axiomatize essentially any theory consisting of a category
equipped with extra structure that is defined algebraically (equationally).
Examples of categorical structures axiomatizable as GATs include categories;
presheaves and copresheaves; monoidal categories, be they strict or weak,
braided or symmetric; categories equipped with chosen finite products or finite
limits; and 2-categories, bicategories, and double categories.

\subsection{Contributions}

In this paper, we describe the design and implementation of
GATlab,\footnote{GATlab is open source software available on GitHub:
  \url{https://github.com/AlgebraicJulia/GATlab.jl}} a programming framework for
generalized algebraic theories built as an embedded domain-specific language in
the Julia programming language. GATs have long served as a foundation for
Catlab,\footnote{Catlab is also available on GitHub:
  \url{https://github.com/AlgebraicJulia/Catlab.jl}} a framework for applied
category theory focusing on scientific and engineering applications. We have
recently rewritten our system for GATs from scratch, while significantly
extending its features. Here we describe it in print for the first time.

Major contributions of GATlab are to
\begin{itemize}
  \item provide an \emph{algebraic specification language} based on a minimal
    dependent type theory in a technical programming language;
  \item support a \emph{standard library} of over 90 reusable theories,\footnote{The
    library of GATs, models, and theory maps can be browsed at:
    \url{https://forest.localcharts.org/gatlib.xml}} ranging from classical
    algebraic structures like groups and rings to categorical structures like
    monoidal categories and presheaves;
  \item enable uniform computing with \emph{models} of GATs, including both
    \emph{free models}, based on symbolic expressions, and computational models,
    defined by arbitrary code in the host language;
  \item declaratively and algebraically \emph{migrate} models of one theory to
    another, via morphisms of theories.
\end{itemize}
In short, GATlab aims to be a structural and symbolic foundation for
categorically structured modeling tools in science and engineering.

\subsection{Related work}

Both the mathematics of GATs and the use of such mathematics in software have a long history. In this section we recount some of that history and explain how GATlab relates to it.

The oldest relevant thread of research is the theory of universal algebra. The phrase ``universal algebra'' dates back at least to Whitehead \cite{whitehead_treatise_1898}, but the subject was not put on a formal footing until Birkhoff's 1946 treatment \cite{birkhoff_Universal_1946}. Later, Lawvere showed in his doctoral thesis how many of the constructions done within universal algebra arise naturally from basic concepts of category theory \cite{lawvere_Functorial_1963}.

Early implementations of universal algebra were in the OBJ and Clear languages \cite{goguen_Abstract_1977,burstall_semantics_1980}. The module system of the Clear language influenced both the design of the Standard ML module system \cite{macqueen_Modules_1984,macqueen_history_2020} and later iterations of the OBJ family of languages \cite{goguen_Introducing_2000}. Modern incarnations of these universal-algebraic concepts can be found in systems such as Maude \cite{clavel_Two_2015}. One of our original objectives for GATlab was to build an ML-inspired module system in Julia, in which implementations of a theory were reifiable objects just as implementations of a signature are in ML. Only later did we realize that ML modules were themselves inspired by universal algebra, so in that sense GATlab draws on a well-established tradition.

GATlab goes beyond that tradition, however, in using generalized algebraic theories, which extend universal algebra to include dependent types, as needed for the theory of categories. The challenge of extending universal algebra with dependent types has been addressed in many ways. One approach is via logical frameworks (LF), first proposed in \cite{harper_framework_1993}. Logical frameworks are an algebraic approach to manipulating the syntax of dependent type theory, by embedding it within another dependent type theory. GATlab plays a similar ``meta'' role as LF, but does not support type constructors which quantify over variables. On the other hand, GATlab offers  other features that LF does not, such as the ability to add arbitrary equations to a theory. That was previously thought to be a bad idea, as it breaks the decidability of type checking and equality checking. However, as we will show, interesting and useful things can be done with GATs even in the absence of decidable type checking.

GATlab's emphasis on theory morphisms as a means of translating between theories and combining theories via colimits is also found in the mathematical knowledge management (MKM) system MMT~\cite{rabe_scalable_2013}, which supports more general classes of theories but is less focused on computational semantics. Likewise, this approach to MKM is found in MathScheme~\cite{carette_MathScheme_2011} and related work on theory presentation combinators~\cite{carette_Theory_2012}.

Finally, GATlab is a computational implementation of categories and other categorical structures. Related projects in this vein include Rydeheard and Burstalls's computational category theory \cite{rydeheard_Computational_1988} and the CAP (Categories, Algorithms, and Programming) computer algebra system \cite{gutsche_Syntax_2018,barakat_CompilerForCAP_2022}.

%% file: 02-background-gats.tex
\section{Background: Generalized algebraic theories and their models}

We review the main ideas behind generalized algebraic theories (GATs), focusing
on their syntax, in both mathematical and GATlab programming notation, and their
standard set-theoretic semantics. GATs were introduced by John Cartmell in his
PhD thesis \cite{cartmell_Generalised_1978} and later in print
\cite{cartmell_Generalised_1986}, to which we refer for the many details omitted
here. As further references, Pitts \cite[\S{6}]{pitts_Categorical_2001} and
Taylor \cite[\mbox{Chapter VIII}]{taylor_Practical_1999} present type theories
based closely on Cartmell's GATs.

\subsection{Syntax of GATs}

GATs are most easily understood through examples. Figure
\ref{fig:theory-categories} shows the ur-example, the theory of categories,
first in mathematical notion in the style of Pitts \cite{pitts_Categorical_2001}
and then as a theory in the GATlab programming environment. Most notably, the
listings are the same up to cosmetic differences in syntax. Julia's macro system
makes this possible by capturing the code in the body of the \texttt{@theory}
macro call instead of evaluating it as standard Julia code. Additionally, GATlab
allows text identifiers to be aliased by mathematical symbols (via Unicode
operators) and axioms to be tagged with names. Note that we break with
convention in type theory and put the context on the right so that the name of
the constructor is the first thing that one sees on a given line.

\begin{figure}
  \centering
  \begin{subfigure}[c]{0.5\textwidth}
    \[ \begin{aligned}
      \Ob : \mathsf{type} &\dashv \\
      \Hom(a,b) : \mathsf{type} &\dashv a, b: \Ob \\
      \\
      f \cdot g : \Hom(a,c) &\dashv a,b,c: \Ob, \\
        & \quad f: \Hom(a,b), g: \Hom(b,c) \\
      \id(a) : \Hom(a,a) &\dashv a: \Ob \\
      \\
      (f \cdot g) \cdot h = f \cdot (g \cdot h) &\dashv a,b,c,d: \Ob, f: \Hom(a,b) \\
        & \quad g: \Hom(b,c), h: \Hom(c,d) \\
      \id(a) \cdot f = f &\dashv a,b: \Ob, f: \Hom(a,b) \\
      f \cdot \id(b) = f &\dashv a,b: \Ob, f: \Hom(a,a)
    \end{aligned} \]
    \subcaption{In mathematical notation}
  \end{subfigure}%
  ~
  \begin{subfigure}[c]{0.55\textwidth}
    \begin{lstlisting}[language=JuliaMin]
@theory ThCategory begin
  Ob::TYPE ⊣ []

  @op (→) := Hom
  Hom(dom::Ob, codom::Ob)::TYPE

  @op (⋅) := compose
  compose(f::(a → b), g::(b → c))::(a → c) ⊣
    [(a,b,c)::Ob]
  id(a::Ob)::(a → a)
  
  assoc := ((f⋅g)⋅h) == (f⋅(g⋅h)) ⊣
    [(a,b,c,d)::Ob, f::(a → b),
     g::(b → c), h::(c → d)]
  idl := id(a) ⋅ f == f ⊣
    [(a,b)::Ob, f::(a → b)]
  idr := f ⋅ id(b) == f ⊣
    [(a,b)::Ob, f::(a → b)]
end
    \end{lstlisting}
    \subcaption{In GATlab notation}
  \end{subfigure}
  \caption{The ur-example of a GAT: the theory of categories}
  \label{fig:theory-categories}
\end{figure}

In general, a GAT is comprised of the following elements.

\begin{description}
  \item[Contexts] A \emph{context} is a finite (ordered) list of named variables
    tagged with types, where the type at a given position may depend on
    variables at previous positions. Examples include the empty context
    \texttt{[]}; the context \texttt{[a::Ob, b::Ob, c::Ob]}, or its shorthand
    \texttt{[(a,b,c)::Ob]}, having no type dependency; and the context
    \texttt{[a::Ob, b::Ob, f::Hom(a,b)]}, where the type of the variable
    \texttt{f} depends on the previous variables \texttt{a} and \texttt{b}. An
    \emph{invalid} context is \texttt{[f::Hom(a,b), a::Ob, b::Ob]} since the
    variables \texttt{a} and \texttt{b} are used as type parameters before they
    are introduced.

  \item[Judgments] A GAT is defined by a sequence of \emph{judgments}. Every
    judgment has a context, which we write to the right of the turnstile
    ($\dashv$). There are three kinds of judgments, summarized in Table
    \ref{tab:gat-judgments}:
    \begin{description}
      \item[Type constructor] New types, possibly dependent on variables in the
        context, are introduced by \emph{type constructors}.

      \item[Term constructor] Typed terms are introduced by \emph{term constructors}.
        Both the term and its type may depend on variables in the context.

      \item[Term equality] Axioms asserting equations between terms in context
        are specified by \emph{term equalities}.
    \end{description}
\end{description}

The theory of categories, for instance, has two type constructors, for objects
and morphisms; two term constructors, for composition and identities; and three
term equalities, for the axioms of associativity and left and right unitality.

As a technical aside, we note that, according to Cartmell, GATs admit a fourth
kind of judgment, equalities between types. We follow Taylor
\cite{taylor_Practical_1999} in excluding type equalities from GATlab. This
restriction simplifies the system, even if it does not solve the problem of type
checking (since dependent types can still be nontrivially equal due to equations
between the terms they depend on). From a category-theoretic perspective,
prohibiting explicit type equalities is no loss since equations between types
are better handled by isomorphisms anyway.

Specifically, this restriction makes sort-checking and sort-inference quite
easy. Sort-inference determines the type constructor used for the type of a
term. For instance, it classifies a term as an object or a morphism, without
necessarily figuring out the domain or codomain of the morphism. This is a
practical and fast check that we perform for most operations in GATlab, and it
catches surface-level bugs.

\begin{table}
  \begin{tabular}{cccp{3in}}
    \toprule
    Judgment kind & Math notation &
      GATlab notation & GATlab examples \\
    \midrule
    Type constructor & $X \;\mathsf{type}\; [\Gamma]$ &
      \texttt{\{X\}::TYPE ⊣ \{Γ\}} &
      \texttt{Ob::TYPE ⊣ []} \newline
      \texttt{Hom(d,c)::TYPE ⊣ [(d,c)::Ob]} \\
    Term constructor & $x : X\; [\Gamma]$ &
      \texttt{\{x\}::\{X\} ⊣ \{Γ\}} &
      \texttt{id(a)::Hom(a,a) ⊣ [a::Ob]} \newline
      \texttt{(a⊗b)::Ob ⊣ [(a,b)::Ob]} \\
    Term equality & $x = y : X\; [\Gamma]$ &
      \texttt{\{x\}==\{y\} ⊣ \{Γ\}} & \texttt{(a⊗b)⊗c == a⊗(b⊗c) ⊣ [(a,b,c)::Ob]} \\
    \bottomrule
  \end{tabular}
  \caption{Judgments in a generalized algebraic theory}
  \label{tab:gat-judgments}
\end{table}

\subsection{Semantics of GATs}

Under the standard set-theoretic semantics, a \emph{model} of a GAT interprets
each type constructor with a set, or family of sets, or family of families of
sets, etc., according to the level of type dependency; and similarly each term
constructor with a function, or family of functions, etc., according to the
level of dependency; in such a way that all of the term equalities are
interpreted as true equations. A more precise account is given by Cartmell
\cite[\S{11}]{cartmell_Generalised_1986}, but the idea becomes most transparent
in examples. A model $M$ of the theory of categories (Figure
\ref{fig:theory-categories}) consists of a set of objects, $M(\Ob)$; a family of
hom-sets, $(M(\Hom)(a,b))_{a,b \in M(\Ob)}$; the identities, a family of
functions $(M(\id)(a): 1 \to \Hom(a,a))_{a \in M(\Ob)}$; and finally the
composition operations, a family of functions
\begin{equation*}
  M(\cdot)(a,b,c): M(\Hom)(a,b) \times M(\Hom)(b,c) \to M(\Hom)(a,c),
  \qquad a,b,c \in M(\Ob),
\end{equation*}
indexed by triples of objects, such that the category axioms are satisfied.
Moving up a dimension, a model $M$ of a theory of 2-categories would include a
family of families of sets $M(\Hom_2)(f,g)$, indexed by $f,g \in M(\Hom)(a,b)$,
indexed in turn by $a,b \in M(\Ob)$.

GATlab offers the programmer two different ways to work with models of GATs, as
we will see. Generalizing their set-theoretic semantics, models of GATs can be
taken in any \emph{contextual category}
\mbox{\cite[\S{14}]{cartmell_Generalised_1986}}. We do not elaborate on the
categorical semantics of GATs as it plays no role in GATlab.

\subsection{Alternatives to GATs}
\label{sec:alternatives-to-gats}

Generalized algebraic theories belong to a family of essentially equivalent
logics that extend the logic of algebraic theories to encompass the theory of
categories and others like it. Besides GATs, the best known of these logics are
essentially algebraic theories \cite[\S{3.D}]{adamek_locally_1994}, finite limit
sketches, and finite limit theories. Cartmell sketched an argument to the effect
that, under their set-theoretic semantics, generalized and essentially algebraic
theories have equivalent categories of models
\cite[\S{6}]{cartmell_Generalised_1986}. Meanwhile, essentially algebraic
theories and finite limit sketches are directly intended to be syntactic
presentations of finite limit theories, an invariant notion of theory.

Any of these logics could, in principle, play the role that GATs do in GATlab.
Our choice of GATs is practical: despite their complicated meta-theory, GATs
lead to by far the most readable and intuitive presentations of theories of
categorical structures, often closely resembling their textbook forms.

%% file: 03-gatlab-gats.tex
\section{Models of GATs in GATlab} \label{sec:gatlab-gats}

In the context of software engineering, GATs play the role of an interface or a formal specification. Models of a GAT, then, are the data structures that implement the interface and satisfy the formal specification.

\subsection{Models of algebraic theories} \label{sec:mlstyle-models}

Given a generalized algebraic theory declared with \texttt{@theory}, we want to be able to ``implement models of that theory in Julia.'' What does this mean?

Languages from the ML family have notions of a \emph{signature} and of a \emph{module} that implements the signature. A module comprises a choice of concrete type for each type in the signature and a function for each term constructor. Models are then referenced explicitly when used. For instance, suppose we had a module \texttt{IntPlusMonoid} implementing the signature of monoids for integers using addition. Then we would use it as follows:
\begin{lstlisting}
  IntPlusMonoid.mappend 1 2 == 3
\end{lstlisting}
We could have another module \texttt{IntTimesMonoid} implementing the signature of monoids for integers using multiplication, and then we'd have
\begin{lstlisting}
  IntTimesMonoid.mappend 1 2 == 2
\end{lstlisting}

The ML module system contrasts with languages like Haskell or Rust, in which type classes or traits have at most a single implementation for a given type, and the implementation is not explicitly referenced. So one must choose between privileging the additive monoidal structure on integers and having \texttt{mappend 1 2 == 3} or privileging the multiplicative monoidal structure on integers and having \texttt{mappend 1 2 == 2}.

In Julia, multiple dispatch can be used to simulate both module-style and trait-style models of theories.\footnote{The original implementation of GATs in Catlab supported only trait-style models, meaning that a given type or collection of types could implement a theory at most one way. A important motivation for creating GATlab was to support both styles of models.}
The trait-style models are most straightforward. For instance, we can implement the free monoid based on string concentation as follows.
\begin{lstlisting}[language=JuliaMin]
  @instance ThMonoid{String} begin
    e() = ""
    (x ⋅ y) = x * y
  end

  "a" ⋅ "b" == "ab" # true
\end{lstlisting}
The \texttt{@instance} macro is a lightweight wrapper around ordinary Julia function definitions using multiple dispatch that checks that all of the methods in the theory have been given implementations. Note, the type parameter syntax (\texttt{\{String\}}) associates concrete Julia types with the GAT's ordered list of type constructors.

In order to implement module-style models using Julia's multiple dispatch, we employ a trick. We associate model implementations with structs (records in Julia), and then use those structs to guide dispatch. In the following code, the \lstinline|Tuple{Int}| parameter to \lstinline{Model} tells the system that when \lstinline{IntPlusMonoid} is used as a model of a theory, the first type constructor in the theory should be interpreted as \lstinline{Int}. We will discuss why we do this later on; in brief this ``exposes'' the implementing types to the Julia type system, which makes it easier to create derived models.
\begin{lstlisting}[language=JuliaMin]
  struct IntPlusMonoid <: Model{Tuple{Int}} end

  @instance ThMonoid{Int} [model::IntPlusMonoid] begin
    e() = 0
    (x ⋅ y) = x + y
  end

  e[IntPlusMonoid()]() == 0 # true
\end{lstlisting}
In the above code, the \texttt{@instance} block is macro-expanded to ordinary Julia code resembling
\begin{lstlisting}[language=JuliaMin]
  ThMonoid.e(::WithModel{IntPlusMonoid}) = 0
  ThMonoid.:(⋅)(::WithModel{IntPlusMonoid}, x::Int, y::Int) = x + y
\end{lstlisting}
The code \texttt{e[IntPlusMonoid()]()} invoked above is syntactic sugar for \texttt{e(WithModel(IntPlusMonoid()))}.

As a technical detail, the abstract type \texttt{Model} takes a tuple type parameter which must match the type parameterization in the \texttt{@instance} declaration, and the type \texttt{WithModel} is a simple wrapper struct defined as:
\begin{lstlisting}[language=JuliaMin]
  struct WithModel{M}
    model::M
  end
\end{lstlisting}
We control multiple dispatch by passing the model type as the first argument; however, we use the wrapped \texttt{WithModel} type to avoid ambiguities that may arise from cases where the instance Julia types are not \texttt{String} or \texttt{Int} as above but rather models themselves, as in the case of the \texttt{SliceC} model, introduced in~\ref{sec:models-with-dependent-types}.

We also a provide a convenience macro \texttt{@withmodel} for evaluating expressions in a module-style model. For example, the invocation
\begin{lstlisting}[language=JuliaMin]
  @withmodel IntPlusMonoid() (e, ⋅) begin
    2 ⋅ e() == 2
  end
\end{lstlisting}
macro-expands to the let block
\begin{lstlisting}[language=JuliaMin]
  let 
    withmod = WithModel(IntPlusMonoid())
    e = () -> ThMonoid.e(withmod)
    ⋅ = (x, y) -> ThMonoid.:(⋅)(withmod, x, y)
    2 ⋅ e() == 2
  end # true
\end{lstlisting}

We are not limited to empty structs when implementing models of theories. For instance, we can implement the monoid $\ZZ/n\ZZ$ of integers modulo $n$ as follows
\begin{lstlisting}[language=JuliaMin]
  struct ModularPlusMonoid <: Model{Tuple{Int}}
    n::Int
  end

  @instance ThMonoid{Int} [model::ModularPlusMonoid] begin
    e() = 0
    (x ⋅ y) = mod(x + y, model.n)
  end
\end{lstlisting}
and use it like so
\begin{lstlisting}[language=JuliaMin]
  @withmodel ModularPlusMonoid(7) (e, ⋅) begin
    3 ⋅ 4 == e()
  end # true
\end{lstlisting}
We call such models \emph{parametric}. They are similar in spirit to ``functors'' in ML languages.\footnote{Incidentally, the name ``parametric modules'' was proposed for functors in ML, but Burstall objected ``on the grounds that we don't call ordinary functions {`parametric values'}'' \cite{macqueen_history_2020}. Since in GATlab they are precisely ``models with parameters,'' we think the name is appropriate.}

\subsection{Incorporating dependent types into models} \label{sec:models-with-dependent-types}

So far we have considered models in GATlab only of algebraic theories. How can we extend these classical notions of model to support dependent types, in light of the fact that Julia doesn't quite have dependent types?\footnote{The question of whether and to what extent Julia has dependent types has engendered much debate \cite{ramine_Julia_2014}. At the very least, we can say that Julia does not have dependent type sufficient for this purpose, as types cannot depend on, say, arbitrary structs. So our implementation of dependent types does not use Julia's dependent types.} There are two approaches to modeling dependent types within a non-dependent language that we have considered. It is illustrative to consider these two approaches in the example of categories.

The first approach is to embrace the translation from generalized algebraic theories to essentially algebraic theories (see Section \ref{sec:alternatives-to-gats}), so that the type constructor $\Hom \colon \Ob \times \Ob \to \Type$ becomes instead a span $\Ob \leftarrow \Hom \to \Ob$ of non-dependent types. Then operations like composition become \emph{partial} functions. We might call this the ``fibered'' approach to dependent types.

In this style, we could implement the category of finite sets, or rather its skeleton, as follows. First, we define data structures for the objects and morphisms of this category.
\begin{lstlisting}[language=JuliaMin]
  struct FinSet
    n::Int
  end

  struct FinFunction
    values::Vector{Int}
    dom::FinSet
    codom::FinSet
  end
\end{lstlisting}
Then we make them into a model of the theory of categories.
\begin{lstlisting}[language=JuliaMin]
  @instance ThCategory{FinSet, FinFunction} begin
    dom(f::FinFunction) = f.dom
    codom(f::FinFunction) = f.codom

    id(x::FinSet) = FinFunction(1:(x.n), x, x)

    function compose(f::FinFunction, g::FinFunction)
      f.codom == g.dom || @fail "domain and codomain do not match"
      FinFunction(g.values[f.values], f.dom, g.codom)
    end
  end
\end{lstlisting}
The \texttt{@fail} macro is a special macro that can be used inside \texttt{@instance} to throw an error with the given message, but also provide contextual information about theory, model, and method in question.

Storing the domain and codomain along with the data of the morphism was the approach taken in the original Catlab implementation of GATs. Indeed, the domain or codomain often cannot be deduced from the data of the morphism. In the example above, representing the data of a function as an array of values determines its range but not its codomain, so the codomain must be stored separately. While intuitive, this approach has the disadvantage that the data structure for a morphism must carry around extra data. So a \emph{diagram} of morphisms might end up storing many redundant copies of the objects.

In GATlab, we continue to support such models, but we also introduce a new approach that cuts down on the redundant storage of objects. The idea is to translate the type constructor $\Hom \colon \Ob \times \Ob \to \Type$ into a predicate $\IsHom \colon \Hom \times \Ob \times \Ob \to \Bool$. Now $\Hom$ is the union of all of the hom-sets, which need not be a disjoint union. If the previous approach was ``fibered,'' we might call this one ``indexed.'' In this style, an implementation of the category of finite sets might look like
\input{finsetc.tex}
Since we can dispatch on the model, we now have the freedom to just use
the types \texttt{Int} and \texttt{Vector\{Int\}} for our objects and morphisms, instead of having to wrap them in custom structs. Note also that in GATlab, the \texttt{Hom} function is not merely a predicate, but rather a coercion function that throws an informative error when the arguments do not type check, and returns a valid morphism when they do.

We can recover the fibered approach from the indexed one by letting $\Hom' = \IsHom^{-1}(\mathbf{true})$, but there is no systematic way to ``drop'' redundant information, so in a sense the second approach is more primitive than the first. This is related to the fact that the Grothendieck construction only gives an equivalence of categories, not an isomorphism.\footnote{One might also think about the indexed approach as a form of the ``type erasure'' that happens during the compilation of dependently typed languages, where values not relevant to the runtime semantics are erased. That is, the runtime representation of values of type \lstinline{Hom(a,b)} need not carry around pointers to \lstinline{a} and \lstinline{b}. Ideally, this would be an optimization performed by a clever compiler rather than done manually, but we must work within the constraints of the Julia programming language.}

We conclude this section by demonstrating how to create a model for slice categories that is parameterized by the base category. First, we declare a model struct.
\begin{lstlisting}[language=JuliaMin]
struct SliceOb{ObT, HomT}
  ob::ObT
  hom::HomT
end

struct SliceC{ObT, HomT, C<:Model{Tuple{ObT, HomT}}}
    <: Model{Tuple{SliceOb{ObT, HomT}, HomT}}
  cat::C
  over::ObT
end
\end{lstlisting}
We interpret this as saying that \texttt{SliceC(cat, over)} is a model with associated types \texttt{SliceOb\{ObT,HomT\}} and \texttt{HomT}. We don't track the fact that it is a model of the theory of categories within the Julia type system because an abstract type in Julia can have only one supertype, but we want to allow for multiple inheritance. We will use the \texttt{over} field as the object within the \texttt{cat} field that we are slicing over. We then implement the methods of the theory of categories as follows.
\begin{lstlisting}[language=JuliaMin]
@instance ThCategory{SliceOb{ObT, HomT}, HomT} [model::SliceC{ObT, HomT, C}]
    where {ObT, HomT, C} begin
  function Ob(x::SliceOb{ObT, HomT})
    try
      Ob[model.cat](x.ob)
    catch e
      @fail ("ob is not valid", e)
    end
    try
      Hom[model.cat](x.hom, x.ob, model.over)
    catch e
      @fail ("hom is not valid", e)
    end
    x
  end

  function Hom(f::HomT, x::SliceOb{ObT, HomT}, y::SliceOb{ObT, HomT})
    try
      Hom[model.cat](f, x.ob, y.ob)
    catch e
      @fail ("morphism is not valid in base category", e)
    end
    compose[model.cat](f, y.hom) == x.hom ||
      @fail "commutativity of triangle does not hold"
    f
  end

  id(x::SliceOb{ObT, HomT}) = id[model.cat](x.ob)

  compose(f::HomT, g::HomT) = compose[model.cat](f, g)
end
\end{lstlisting}
The most interesting parts are in the dynamic checks for validity of objects and morphisms. In the function \texttt{Ob}, we check that \texttt{x.ob} is a valid object inside \texttt{model.cat}, and then we check that \texttt{x.hom} is a valid morphism from \texttt{x.ob} to \texttt{model.over}. In the function \texttt{Hom}, we check that \texttt{f} is a valid morphism from \texttt{x.ob} to \texttt{y.ob} and that the relevant triangle commutes. The last two methods, \texttt{id} and \texttt{compose}, are just inherited from the base category \texttt{model.cat}.

\subsection{Free models of GATs}

In addition to models based on arbitrary Julia code, GATlab can construct \emph{free} or \emph{symbolic} models of GATs,\footnote{Strictly speaking, these ``free models'' are not necessarily models of the theory, because the theory's equational axioms may hold only up to rewrites of the expressions.} via the \texttt{@symbolic\_model} macro. For instance, the following macro call creates a free model of theory of categories:
\begin{lstlisting}[language=JuliaMin]
  @symbolic_model FreeCategory{ObExpr, HomExpr} ThCategory begin
    compose(f::Hom, g::Hom) = associate_unit(new(f,g; strict=true), id)
  end
\end{lstlisting}
This code creates a new module called \texttt{FreeCategory} in which structs called \texttt{Ob} and \texttt{Hom} are defined and methods \texttt{ThCategory.id} and \texttt{ThCategory.compose} implemented for them. It also provides some special logic via an externally-defined function \lstinline{associate_unit} to enforce that composition is strictly associative and unital, by normalizing to $n$-ary composition operations whenever \lstinline{compose} is called. We can use the resulting data structures as follows.
\begin{lstlisting}[language=JuliaMin]
  A, B = Ob(FreeCategory, :A), Ob(FreeCategory, :B)
  f, g = Hom(:f, A, B), Hom(:g, B, A)
\end{lstlisting}
Essentially, we can think of this model as ``the free model of the theory of categories on the set of symbols.'' Specifically, this is the free category on the graph whose vertices are the set of symbols and whose edges between any two vertices are also the set of symbols.

One use of free models in Catlab is to schedule and evaluate string diagrams. That is, we can transform a combinatorial presentation of a string diagram into an expression in the theory of symmetric monoidal categories. If we then have a model of the theory of symmetric monoidal categories, we can then interpret this expression in that model.

Within the \texttt{@symbolic\_model} block, we can specialize the implementation of the operations of the theory to normalize the terms. For instance, in the invocation of \texttt{@symbolic\_model} above, we normalize with respect to the associative law and the unit law, so that \texttt{compose(f,compose(g,h)) == compose(compose(f,g),h)}. Normalization isn't always possible for more complex theories, but it is convenient when it is possible.

Currently, we have implemented the following normalization strategies:

\begin{itemize}
\item \lstinline{associate}, normalizing a binary operation for associativity
\item \lstinline{associate_unit}, normalizing a binary operation for associativity and unitality
\item \lstinline{associate_unit_inv}, normalizing a binary operation for associativity, unitality, and inverses
\item \lstinline{distribute_unary}, distribute a unary operation over a binary operation
\item \lstinline{involute}, normalize an involutive unary operation
\item \lstinline{normalize_zero}, collapse to zero when the expression contains a zero
\end{itemize}

However, these are simply the normalization strategies that we have found useful; our approach does not preclude other strategies from being implemented.

%% file: finsetc.tex
\begin{lstlisting}[language=JuliaMin]
  struct FinSetC <: Model{Tuple{Int, Vector{Int}}} end

  @instance ThCategory{Int, Vector{Int}} [model::FinSetC] begin
    Ob(x::Int) = x >= 0 ? x : @fail "expected nonnegative integer"

    function Hom(f::Vector{Int}, n::Int, m::Int)
      length(f) == n ||
        @fail "length of morphism does not match domain: $(length(f)) != $n"
      for i in 1:n
        if !(f[i] in 1:m)
          @fail "index not in codomain: $i"
        end
      end
      return f
    end

    id(x::Int) = collect(1:x)
    compose(f::Vector{Int}, g::Vector{Int}) = g[f]
  end
\end{lstlisting}

%% file: 04-background-gat-morphisms.tex
\section{Morphisms of GATs}

While it is always possible to write arbitrary functions that transform a model
of one theory into a model of another, the code that one must write is often
opaque and not easily verified. In many situations, a solution to this problem
is provided by the fact that different GATs are related to each other by
\emph{morphisms}, also called \emph{interpretations}. Using morphisms of GAT, we
implement a declarative and verifiable syntax for migrating models from one
theory to another.

\subsection{Syntax of GAT morphisms}

Following Cartmell \cite[\S{12}]{cartmell_Generalised_1986}, two GATs $A$ and
$B$ can be related to each other via an {\em interpretation}, which assigns to
each type (respectively, term) constructor of $A$ a type (respectively, term) in
$B$. This mapping must satisfy some conditions in order for it to be a
well-typed and validity-preserving translation of the vocabulary of $A$ into the
vocabulary of $B$.

For example, consider translating statements in the language
of monoids into the language of natural number arithmetic:
\begin{center}
\begin{minipage}{.45\textwidth}
\begin{lstlisting}[language=JuliaMin]
  @theory ThMonoid <: ThSet begin
    (x⋅y) :: default ⊣ [x, y]
    (x⋅y)⋅z == (x⋅(y⋅z)) ⊣ [x, y, z]
    e() :: default
    e()⋅x == x ⊣ [x]
    x⋅e() == x ⊣ [x]
  end
\end{lstlisting}\end{minipage}%
\hspace{1pt}
\begin{minipage}{0.49\textwidth}
\begin{lstlisting}[mathescape, language=JuliaMin]
  @theory ThArith <: ThNat begin 
    (x::$\mathbb{N}$ + y::$\mathbb{N}$)::$\mathbb{N}$ ⊣ [(x, y)::$\mathbb{N}$]
    n + S(m) == S(n+m) ⊣ [(n, m)::$\mathbb{N}$]

    (x::$\mathbb{N}$ * y::$\mathbb{N}$)::$\mathbb{N}$
    n*S(m) == ((n*m) + n) ⊣ [(n,m)::$\mathbb{N}$]
  end
\end{lstlisting}\end{minipage}
\end{center}
Note that \lstinline{ThSet} is a special theory with one type,
\lstinline{default}, which is the default interpretation of a type if the user
does not provide one. We can interpret the theory of monoids either additivity
or multiplicatively in the theory of arithmetic:
\begin{center}
\begin{minipage}{.4\textwidth}
\begin{lstlisting}[mathescape, language=JuliaMin]
  @map PlusM(ThMonoid, ThArith) begin
    default => $\mathbb{N}$
    x⋅y ⊣ [x, y] => x+y
    e() => Z()
  end
\end{lstlisting}\end{minipage}%
\hspace{30pt}
\begin{minipage}{0.45\textwidth}
\begin{lstlisting}[mathescape, language=JuliaMin]
  @map TimesM(ThMonoid, ThArith) begin  
    default => $\mathbb{N}$
    x⋅y ⊣ [x, y] => x*y
    e() => S(Z())
  end
\end{lstlisting}\end{minipage}
\end{center}

In general, to declare how an interpretation $I$ acts on a type (resp. term)
constructor $f$, on the left side of the arrow we write $f$'s {\it generic} type
(resp. term) in context, ${f(x_1,...,x_n) \dashv [arg_1::T_1,...,arg_n::T_n]}$. On the right side we define
$I(f)$ by writing a type (resp. term) implicitly in the context
${[x_1::I(T_1),...,x_n::I(T_n)]}$. This mapping defined on type and term
constructors induces a function on all type and term expressions using the
operations of $A$ into expressions of $B$; we say that this interpretation
function ``pushes forward types/terms/contexts along $I$". For example, the
expression
\lstinline{PlusM(e()⋅x⋅e() ⊣ [x])}
evaluates to
\lstinline{Z()+x+Z() ⊣ [x::ℕ]}.
We call an interpretation $A \rightarrow B$ a \emph{theory map} or
\emph{morphism} in the category $\mathsf{GAT}$.

The above examples can be contrasted with user-declared morphisms that are
ill-typed or give invalid interpretations:
\begin{center}
\begin{minipage}{.40\textwidth}
\begin{lstlisting}[mathescape, language=JuliaMin]
@map Bad₁(ThMonoid, ThArith) begin
  default => $\mathbb{N}$
  x⋅y ⊣ [x,y] => x+y
  # unit axioms invalidated
  e() => S(Z())
end
\end{lstlisting}\end{minipage}%
\hspace{8pt}
\begin{minipage}{0.55\textwidth}
\begin{lstlisting}[language=JuliaMin]
@map Bad₂(ThCategory, ThCategory) begin
  # Ill-typed arguments for Hom
  Ob ⊣ [] => Hom()
  a → b ⊣ [(a,b)::Ob] => a → b
  # Ill-typed argument for id
  id(a) ⊣ [a::Ob] => id(a)
end
\end{lstlisting}\end{minipage}
\end{center}
An interpretation is ``validity-preserving" when, for each axiom $Eq$ of $A$,
there exists a proof of $I(Eq)$ using the axioms of $B$. When $I$ is
\lstinline{Bad₁} and $Eq$ is the unitality axiom $e()\cdot x = x$, this would mean
that we need to find a proof that $1+x = x$ in \lstinline{ThArith}, which is
impossible. The other possible issue with a user-specified theory morphism is
that it is ill-typed, such as \lstinline{Bad₂}, which attempts to map
$Ob \mapsto Hom()$, which fails both because \lstinline{Hom} requires arguments and
because \lstinline{ThCategory} requires that \lstinline{id} take an
\lstinline{Ob} argument, not a \lstinline{Hom}.

\subsection{Data structures for subclasses of GAT morphisms}

Morphisms of GATs are highly expressive, which entails a certain amount of
complication. It can be useful to consider GAT morphisms that are more
restrictive, yet require less data to specify. These more restricted GAT
morphisms also make it easier to compute, for instance, pushouts.  GATlab
defines no less than four data structures that implement GAT morphisms of
varying expressivity:
\begin{center}
\begin{minipage}{.4\textwidth}
\begin{lstlisting}[language=JuliaMin]
struct IdTheoryMap <: AbsTheoryMap
  gat::GAT
end

struct SimpleTheoryMap <: AbsTheoryMap
  dom::GAT; codom::GAT
  typemap::Dict{Ident, Ident}
  termmap::Dict{Ident, Ident}
end
\end{lstlisting}\end{minipage}%
\hspace{35pt}
\begin{minipage}{0.4\textwidth}
\begin{lstlisting}[language=JuliaMin]
struct InclTheoryMap <: AbsTheoryMap
  dom::GAT; codom::GAT
end

struct TheoryMap <: AbsTheoryMap
  dom::GAT; codom::GAT
  typemap::Dict{Ident, TypeInCtx}
  termmap::Dict{Ident, TermInCtx}
end
\end{lstlisting}\end{minipage}
\end{center}
Each of these kinds of morphism can be cast to the subsequent kind, thus all can
be interpreted as general theory maps, which send type and term constructors
respectively to types and terms in a context of the codomain theory.
\lstinline{IdTheoryMap} represents the identity morphism on a GAT, and
\lstinline{InclTheoryMap} represents inclusions of theories. The third kind,
\lstinline{SimpleTheoryMap}, allows for associating type and term constructors
of the domain with structurally-identical type and term constructors of the
codomain. The fourth and final kind, \lstinline{TheoryMap}, is a general
morphism of theories, as described above.

For example, consider a map \lstinline{F} which interprets preorders
in the language of categories:
\begin{center}
\begin{minipage}{.54\textwidth}
\begin{lstlisting}[language=JuliaMin]
@theory ThPreorder <: ThSet begin
  Leq(dom, codom)::TYPE ⊣ [dom, codom]
  @op (≤) := Leq
  refl(p)::(p ≤ p) ⊣ [p]
  tran(f::p ≤ q, g::q ≤ r)::(p ≤ r) ⊣ [p,q,r]
  irrev := f == g ⊣ [p, q, (f, g)::(p ≤ q)]
end
\end{lstlisting}
\end{minipage}%
\hspace{-5pt}
\begin{minipage}{0.4\textwidth}
\begin{lstlisting}[language=JuliaMin]
@map F(ThCategory, ThPreorder) begin
  Ob => default
  Hom => Leq
  compose => tran
  id => refl
end
\end{lstlisting}
\end{minipage}
\end{center}
There is no overlap between the domain and codomain GATs, so this map cannot be
represented by an \lstinline{InclTheoryMap}, but it is a
\lstinline{SimpleTheoryMap}. In addition, \lstinline{PlusM} and
\lstinline{TimesM} above could both have been expressed using
\lstinline{SimpleTheoryMap} syntax. However, a general \lstinline{TheoryMap} is
needed to represent the following GAT morphism, which takes a monoid to its
opposite:
\begin{lstlisting}[language=JuliaMin]
  @map OpMonoid(ThMonoid, ThMonoid) begin
    default => default;  e() => e();  x⋅y ⊣ [x, y] => y⋅x
  end
\end{lstlisting}
When \lstinline{OpMonoid} is applied to \lstinline{x⋅(e()⋅y) ⊣ [x, y]} one
obtains \lstinline{(y⋅e())⋅x ⊣ [x, y]}.

\subsection{Computation with GAT morphisms}

The basic functionalities these data structures must provide in order to play
the role of GAT morphisms are composition, pushing forward of terms, pulling
back of models, and checking the validity of the morphism. All of these
operations are trivial for morphisms of kind \lstinline{IdTheoryMap} or
\lstinline{InclTheoryMap}. Composition for \lstinline{SimpleTheoryMap} is simply
composition of the underlying functions from identifiers to identifiers, whereas
composition for \lstinline{TheoryMap} is implemented as a composition of
substitutions, akin to a Kleisli composition for a term monad.

Pushing forward types and terms along \lstinline{SimpleTheoryMap} or
\lstinline{TheoryMap} instances requires an implementation of pattern matching
and syntactic substitution. The pushforward operation is essential to pulling
back models along theory maps: models associate a computational semantics with
each expression of a theory $B$, so by precomposing with the interpretation
function of expressions of $A$ to $B$, we see that GAT
morphisms $A \rightarrow B$ induce a function from models of $B$ to models of
$A$. For example, if \lstinline{ThArith} were given a model
\lstinline{JuliaArith}, using the native addition and multiplication of Julia,
then the model \lstinline{IntPlusMonoid} defined manually in the previous
section could be obtained declaratively by calling GATlab's
\lstinline{migrate_model} on \lstinline{PlusM} and \lstinline{JuliaArith}.
As another example, the following theory map allows us to automatically generate
models of opposite categories. Once we have checked that the theory map is valid,
we have confidence that applying it to correct models will always produce correct
models.
\begin{lstlisting}[language=JuliaMin]
  @map OpCat(ThCategory, ThCategory) begin
    Ob => Ob
    (dom → codom) ⊣ [(dom, codom)::Ob] => (codom → dom)
    id(a) ⊣ [a::Ob] => id(a)
    compose(f, g) ⊣ [(a,b,c)::Ob, f::(a → b), g::(b → c)] => compose(g, f)
  end
\end{lstlisting}

The operation of checking the validity of a theory map is trivial for
\lstinline{IdTheoryMap}. For \lstinline{InclTheoryMap}, one must check that the
identifiers found in \lstinline{dom} are a strict subset of those found in
\lstinline{codom}. For \lstinline{SimpleTheoryMap}, one must check that for each
type and term constructor of the domain that the assigned type or term in the
codomain has the same structure. E.g., to check the map \lstinline{F}'s assignment
\lstinline{compose => tran},
one looks at the domain's generic term in context, namely
\begin{lstlisting}[language=JuliaMin]
  compose(f,g) ⊣ [(a,b,c)::Ob, f::(a → b), g::(b → c)]
\end{lstlisting}
and checks whether applying the theory map to the term in context is identical
(up to renaming variables) to the generic term in context of the codomain
theory, which in this case is
\begin{lstlisting}[language=JuliaMin]
  tran(f, g) ⊣ [(a,b,c)::default, p::(a≤b),q::(b≤c)]
\end{lstlisting}
For a general theory map, determining validity-preservation requires solving
many word problems, and deciding equality is a nontrivial semi-decidable
computation in general. For example, one must check that \lstinline{OpMonoid}
preserves the associativity of monoid multiplication.

Furthermore, morphisms can be used to compositionally construct new theories via
colimits. This can be done without explicit morphisms due to special support for
a multiple inheritance syntax:
\begin{lstlisting}[language=JuliaMin]
  @theory ThModule begin
    using ThRing: default as Scalar
    using ThAdditiveAbelianGroup: default as Vector
    α ⋅ v :: Vector ⊣ [α::Scalar, v::Vector]
    # ... axioms
  end
\end{lstlisting}
\lstinline{ThModule} is an extension of the coproduct of \lstinline{ThRing} and \lstinline{ThAdditiveAbelianGroup}.

Alternatively, the same theory can be explicitly constructed as a pushout of
maps to suitably renamed versions of \lstinline{ThRing} and
\lstinline{ThAdditiveAbelianGroup} from the theory of their common overlap (in
this case, the empty theory), which can then be extended to include scalar
multiplication and axioms. When the span to be pushed out uses only
\lstinline{SimpleTheoryMap} or simpler kinds of morphism, the computation is a
simple matter of renaming and taking the union of the component GATs. Pushouts
of general theory maps are a much more complicated matter yet to be implemented.
Thus, drawing distinctions between different flavors of theory map allows GATlab
to distinguish feasible and infeasible computations with GATs.

%% file: 05-applications.tex
\section{Applications and Extensions}

In creating the GATlab package as a replacement for the GAT machinery in Catlab, our first major goal was to explore new design decisions such as
\begin{enumerate}
  \item models of GATs in the ML module style, in addition to the Haskell typeclass style (\cref{sec:mlstyle-models})
  \item fibered vs. indexed point of view on dependent types (\cref{sec:models-with-dependent-types})
  \item hygenic substitution via scope tags (\cref{sec:scopes})
\end{enumerate}
but retain sufficient backwards-compatibility with Catlab that adoption of these new features could be done incrementally rather than all at once. Having now completed this significant engineering effort, we are beginning to take advantage of GATlab's new functionality throughout the AlgebraicJulia ecosystem of packages for applied category theory.

One of the motivations behind GATlab was to build functionality for symbolic dynamical systems. In the past, we have built dynamical systems on top of arbitrary Julia functions \cite{libkind_Operadic_2022}. However, making the various functions in dynamical systems \emph{symbolic} unlocks a range of new capabilities, such as:

\begin{itemize}
  \item Systems can be optimized/compiled to produce more efficient code
  \item Systems can be analyzed symbolically to prove properties about them without simulating them
  \item Systems can be transferred between different programming languages
  \item The equations defining a system can be pretty-printed and displayed to the user
\end{itemize}

There is a blog post written by the first author which explores these directions \cite{lynch_Symbolic_2023}, and we have a preliminary implementation of symbolic resource sharers as an operad algebra \cite{lynch_Symbolic_2024}. Notably, GATlab allows us to implement these symbolic resource sharers agnostic to the exact collection of ``primitive functions'' usable in the implementation of a symbolic function. For instance, we could only allow polynomial functions, or only allow polynomial and trigonometric functions, and so on.

To support these systems-theoretic capabilities, another major direction for GATlab is developing richer functionality for computer algebra. We are interested in exploring how types can aid the development of computer algebra for mathematical objects going beyond classical abstract algebra. For instance, we have prototyped a theory in which the primary objects of study are ``sections of vector bundles,'' where both the vector bundle itself and the space over which the vector bundle sits are arguments to the type constructor. This would enable principled, coordinate-free descriptions of problems in differential geometry, and have applications to partial differential equations via use in Decapodes \cite{patterson_diagrammatic_2022,morris_Decapodes_2024}.

There are several ways in which we are considering adding more sophisticated rewrite functionality to GATlab for the purposes of computer algebra. We have prototyped our own implementation of e-graphs for GATlab syntax trees, based on a careful reading of the egg source code \cite{willsey_egg_2021}. However, Julia already has an excellent implementation of e-graphs in the form of Metatheory.jl \cite{cheli_Metatheory_2021}, with which we hope to integrate. In any case, the advantage of e-graphs is that they are applicable to almost any rewriting problem, which is a good fit for the generality of GATs. However, in specialized domains like rings and modules, algorithms that involve Groebner bases can be far more efficient than naive use of e-graphs. We have access to such computer algebra systems in Julia via Symbolics.jl \cite{gowda_High_2022} and the CAP project \cite{barakat_CompilerForCAP_2022}, and so another direction would be to integrate these into specific theories within GATlab.

A limitation of GATs is that although we can specify a GAT for categories, GATs only have a 1-category of models rather than a 2-category. So, from the GAT for categories, we can automatically derive the concept of functor, but not the concept of natural transformation. Lambert and Patterson have a notion of ``cartesian double theory'' \cite{lambert_Cartesian_2024}, the models of which naturally form a 2-category, which could be a more natural framework for doing ``computer algebra for categories'' than generalized algebraic theories.

Yet another direction for future research is randomized testing. This can be done in two ways. The first is to take an arbitrary equation in a theory, the validity of which is uncertain, and then check if it is true in randomly sampled finite models of a GAT. Of course, this cannot prove that the equation is true, but it can furnish counterexamples in the style of Nitpick \cite{blanchette_Nitpick_2010} and Alloy \cite{milicevic_Alloy_2015}. The second way is to take a purported model of a GAT, and then sample elements of the types in that model to check if the equations in the GAT actually hold, in the style of QuickCheck \cite{claessen_QuickCheck_2000} and Hypothesis \cite{maciver_Hypothesis_2019}. This would provide a cheap way to get confidence in an implementation, and improve the overall code quality of anything which used GATlab.

%% file: A1-implementation.tex
\section{Implementation of GATlab}

In this section, we describe certain details of GATlab's implementation that may
be useful to developers looking to create similar systems.

\subsection{Scopes} \label{sec:scopes}

In the original Catlab implementation of GATs, variables were identified simply by symbols, which implicitly means that variables were resolved via dynamic scoping. An objective of the GATlab implementation was to provide a lexical scope system, which properly handled issues like shadowing and overloading and helped identify errors in syntax manipulation algorithms.
We originally intended to use a system based on deBruijn levels, but the reindexing necessary when moving syntax around while using deBruijn levels proved quite tedious to handle. We eventually settled on a system inspired by Racket's ``scope sets'' construct, which is used to ensure hygenicity of the Racket macro system \cite{flatt_Bindings_2015}. However, because GATlab does not involve macros that expand to macros, we did not need the full generality of ``scope sets.''

Our scope system works in the following way. We call any piece of syntax that introduces new variable bindings a \textbf{scope}. For instance, the body of a \texttt{@theory} macro is a scope, the right hand side of a turnstile (``$\dashv$'') is a scope, and the collection of arguments to a function is a scope.

At parse time, every scope is assigned a random UUID, which we call a \textbf{scope tag}. An \textbf{identifier} consists of a scope tag, an index ranging from 1 to the length of the scope, and a name. The scope tag and the index are sufficient to identify the binding that the identifier refers to; the name is just for display purposes.

A context is then an ordered list of scopes. When a symbol is parsed within a context, we produce an identifier with a scope tag that refers to the latest scope within that context that has a binding named with that symbol. However, when later on we want to find the binding associated with an identifier, we look through the context for a scope with a matching scope tag, rather than simply taking the first binding with the right name. The scope mechanism ensures that substitution is \emph{hygenic}: when a variable is substituted into a new expression, in a new environment, it will never accidentally refer to something it was not intended to refer to.

While implementing lexical scopes might seem a minor detail, it was not trivial to settle on this approach to scoping, and so we hope that reporting the details of this solution will help designers of typed computer algebra systems in the future. Using a combination of UUID-based scope identifiers and consecutive-integer indexing is more ergonomic than pure deBruijn indexing because reindexing happens less frequently. More importantly, it is also better at detecting programmer errors. In a computer algebra system, we are frequently moving syntax trees around, so it may happen that we forget to bring the corresponding scope along; having variables reference their scope by UUIDs is an easy way to detect this kind of error.

\subsection{Expanding macros inside macros}

This section is about a trick somewhat specific to Julia as we do not know how many other programming languages support calling \texttt{macroexpand} inside a macro. It would be an interesting exercise to make this work in a language like Racket which has strict phase separation.

When one theory in GATlab extends another theory, we need to access the contents of the parent theory \emph{at macro expansion time}. The reason is that the \texttt{@theory} macro produces Julia modules and declares functions, so if we had to wait until runtime to get the contents of the parent theory, we would have to call \texttt{eval} at runtime to produce those modules. Exactly why this would not be ideal is a complicated story involving the details of how Julia performs compilation, but suffice to say that this is best avoided.

Generally, macros cannot access the \emph{values} referred to by the syntax that is passed to them, only the syntax itself. However, one thing that \emph{can} be done inside a Julia macro is to call \texttt{macroexpand}. We exploit this possibility in the following way. A call to \texttt{@theory ThMonoid begin ... end} expands into a module called \texttt{ThMonoid} as follows.

\begin{lstlisting}[language=JuliaMin]
  module ThMonoid
  ...
  module Meta
    ...
    macro theory()
      # (return the data of the theory of monoids)
    end
    ...
  end
  ...
  end
\end{lstlisting}
Then, when macroexpanding an extension of the theory
\begin{lstlisting}[language=JuliaMin]
  @theory ThGroup <: ThMonoid begin
    ...
  end
\end{lstlisting}
we do something like
\input{theorymacro.tex}

We now have access to the data of the theory of monoids in the variable \texttt{parenttheory}, and can use it in the rest of the \texttt{@theory} macro.

%% file: theorymacro.tex
\begin{lstlisting}[language=JuliaMin]
  macro theory(head, body)
    parentname = ...
    parenttheory = macroexpand(__module__, :($parentname.Meta.@theory))
    ...
  end
\end{lstlisting}